\documentclass[conference]{IEEEtran}
\IEEEoverridecommandlockouts
\usepackage{cite}
\usepackage{amsmath,amssymb,amsfonts}
\usepackage{algorithmic}
\usepackage{graphicx}
\usepackage{textcomp}
\usepackage{xcolor}
\usepackage{colortbl}
\usepackage{iftex}
\ifXeTeX
  \usepackage{fontspec}
\else
  \usepackage[T1]{fontenc}
\fi
\usepackage{booktabs}
\usepackage{url}
\newcommand{\cmark}{$\checkmark$}
\newcommand{\xmark}{$\times$}
\newcommand{\pmark}{$\triangle$}

\def\BibTeX{{\rm B\kern-.05em{\sc i\kern-.025em b}\kern-.08em
    T\kern-.1667em\lower.7ex\hbox{E}\kern-.125emX}}

\usepackage{eso-pic}
\newcommand{\ieeecopyrightnotice}{%
  \AddToShipoutPictureBG*{%
    \AtPageLowerLeft{%
      \put(\LenToUnit{0.5\paperwidth},\LenToUnit{0.32in}){%
        \makebox[0pt][c]{\parbox{\textwidth}{\centering\scriptsize
        \copyright\ 2026 IEEE. Personal use of this material is permitted. Permission from IEEE must be obtained for all other uses, in any current or future media, including reprinting/republishing this material for advertising or promotional purposes, creating new collective works, for resale or redistribution to servers or lists, or reuse of any copyrighted component of this work in other works.}}}}}}
\begin{document}

\title{Beyond Link Quality: Congestion-Aware Sequential Offloading of Robot VLM Inference over Private 5G\\
\thanks{This work was supported by Institute of Information \& communications Technology Planning \& Evaluation (IITP) grant funded by the Korea government (MSIT) (No. RS-2025-25457382, Development of Operation and Deployment Technologies for Korean AI SoC based Micro Data Centers)}
}

\author{\IEEEauthorblockN{Hyeju Shin, Jusung Kang, Gicheol Wang, Ryangsoo Kim, Jaein Kim\textsuperscript{*}, Hark Yoo\textsuperscript{*}%
\thanks{\textsuperscript{*}Corresponding author: Jaein Kim (jaein@etri.re.kr), Hark Yoo (harkyoo@etri.re.kr)}}
\IEEEauthorblockA{\textit{Edge Computing Application Service Research Section} \\
\textit{Electronics and Telecommunications Research Institute (ETRI)}\\
Gwangju, South Korea \\
\{hyeju.shin, k92492, gcwang, rskim, jaein, harkyoo\}@etri.re.kr}
}

\maketitle
\ieeecopyrightnotice

\begin{abstract}
Robots need vision-language models (VLMs) to describe and evaluate their surroundings. However, VLM inference on embedded devices is slow and energy-intensive, which motivates offloading to private 5G multi-access edge computing (MEC) servers and micro data centers (MDCs). Existing policies weigh energy, latency, and the link for a single task, so they miss that model compute decides whether link degradation or server load matters, and that each choice changes the modem's radio resource control (RRC) state and later costs. Measurements of object detection, pose estimation, and tiny and small VLMs in three power modes on a real 5G standalone (SA) MEC testbed show that, without server load, requests fall into local-favored, state-dependent, and offload-favored regimes. Link degradation changes decisions only in the state-dependent regime, whereas server load does so even for the small VLM. We formulate a constrained Markov decision process (MDP) in which the fixed violation penalty acts as a Lagrange multiplier. We then show that a myopic policy without exploration can become trapped in local execution under RRC power-saving. Our PPO-based congestion-aware sequential offloading policy observes the link metrics, task compute, power mode, RRC state, and server load. Its reward combines board energy, per-type latency violations, and LLM-judged answer quality. In simulated replays of field drives, it cuts the violation rate under server load from 40.4\% (SW-UCB) to 23.6\%. On the real device in MAXN, where the local model meets the constraints, it cuts the violation rate from 47.1\% to 1.4\% (mean over four seeds) at slightly lower long-question quality. Offloading VLMs to shared edge servers therefore requires observing the model compute, power mode, and server load together with the link and the modem state.
\end{abstract}


\section{Introduction}\label{sec:intro}

Physical AI robots that work alongside people are expected to describe and judge their surroundings in addition to detecting obstacles~\cite{r2}, for example by answering questions such as ``Is it safe to go straight now, and why?'' with vision-language models (VLMs)~\cite{r3}. However, embedded devices run these models slowly. On a Jetson AGX Orin in its highest-performance mode, a small VLM (Qwen2.5-VL-3B~\cite{r6}) takes about 1.5s and 77J per question. In the 15W mode, it takes over 8s, whereas object detection takes about 20ms. Although quantization helps with memory, it can slow down processing on Jetson devices~\cite{r58}. Therefore, offloading to an edge server is a natural solution~\cite{r7,r15}, especially on private 5G networks~\cite{r9}, where the multi-access edge computing (MEC) server is placed next to the core network, as in AI-RAN~\cite{r10}. However, offloading is not always beneficial. When the link degrades or the shared server is congested, latency increases, and the device continues to draw power while waiting for a response. A policy must therefore decide on a per-request basis whether to run locally or offload.

Previous offloading work considered energy and the link, along with accuracy~\cite{r13} or computation rate~\cite{r14}, and some DRL-based offloading has been evaluated on real devices over 5G SA~\cite{r57}. These works, however, model server load only as a device task queue or available server resources and do not model the radio resource control (RRC) state of the modem (Table~\ref{tab:related}). Kim et al.~\cite{r15} reduced per-frame energy consumption under a latency constraint by learning the costs of both options per RSSI bin using SW-UCB~\cite{r16} over indoor Wi-Fi. However, with one vision model, signal strength as the only state, and independent decisions, it can observe neither server load nor how a choice affects future costs.

We revisited these assumptions on a real 5G standalone (SA) private network with an MEC server connected to the user plane function (UPF). The round-trip time for a vision request on this network is only about 16ms. The offloading gain is set mainly by the model compute, and the link changes the decision only for some requests (Fig.~\ref{fig:compute}). Lightweight object detection runs faster locally, whereas the remote 7B VLM takes only 0.55$\times$ the latency and 0.17$\times$ the energy of the local 3B. For pose estimation and short VLM questions, the decision depends on the device power mode, the link, the server load, and the latency constraint of each request.

This reveals three limitations of existing policies. First, distinguishing states only by signal strength pools the costs of different tasks in one bin, which reduces the reward at the cell edge (Section~\ref{sec:s2}). Second, treating requests independently ignores the fact that a series of local executions puts the modem into an RRC power-saving state, which delays the next offload~\cite{r17}. Third, without observing server load, such policies continue to offload into a congested queue and violate latency constraints (Section~\ref{sec:s3s4}). We therefore propose a congestion-aware sequential offloading policy, where congestion refers to server load from other clients rather than network congestion. This policy adds the task's compute, the device's power mode, the RRC state, and the server load to the 5G link metrics. Its reward combines device energy, violations of per-question type latency constraints, and answer quality. Since the action changes the RRC state, we formulate a Markov decision process (MDP) and learn the policy with PPO~\cite{r18}; with the quality weight set to zero and a single task, the objective reduces to that of~\cite{r15}, which allows a fair comparison. Our contributions are:
\begin{itemize}
\item \textbf{Measurement on private 5G MEC.} Without server load, four tasks (OD, PE, tiny VLM, small VLM) under three power modes fall into \emph{local-favored}, \emph{state-dependent}, and \emph{offload-favored} regimes; link degradation matters only in the state-dependent regime, whereas server load matters even for the offload-favored small VLM.
\item \textbf{Congestion-aware sequential offloading.} A constrained-MDP formulation that interprets the fixed violation penalty as a Lagrange multiplier, and a proposition that a greedy myopic policy without exploration can lock into local execution under RRC dynamics. The gain over link-only policies comes from observing the compute, power mode, RRC state, and server load together with the link, and a load-threshold rule tuned per task and power mode on the same state performs about as well. In the ablation, the load mainly saves energy and the RRC state reduces violations under load.
\item \textbf{One policy for multiple models, validated on a real device.} One policy with per-task compute as input handles the four tasks and, without retraining, serves an unseen tiny VLM (SmolVLM2) that shares the remote model, with an S4 reward no lower than that of a policy trained with it. It cuts the violation rate from 40.4\% (SW-UCB) to 23.6\% in simulation of the server-load scenario (S4), and from 47.1\% to 1.4\% on the real device in MAXN, where the local model meets the latency constraint (33.2\% vs.\ 56.6\% averaged over the three power modes). Its S4 long-question quality is lower (0.405 vs.\ 0.424 for SW-UCB), and its S1 reward is slightly below Always-Remote's.
\end{itemize}

\section{Related Work}\label{sec:related}

\textbf{Offloading decisions.} Bandit approaches adapt online; our baseline~\cite{r15} learns local and remote costs per RSSI bin with SW-UCB~\cite{r16} over indoor Wi-Fi with a penalty for latency violations, and we inherit its problem setting and cost. Deep RL approaches include DROO~\cite{r14}, deadline-aware dueling double DQN~\cite{r22}, and PPO for UAV-assisted MEC~\cite{r24}; PPO~\cite{r18} has been reported to train more stably than other policy-gradient and DQN methods in offloading with task queues~\cite{r55} and DNN inference~\cite{r56}. Closest to our work on real hardware, Nieto et al.~\cite{r57} let devices including a Jetson AGX Orin choose among local, MEC, and cloud execution with DRL over 5G SA and measure energy, but with synthetic tasks and only throughput as the radio state, without the RRC state or the server load. These works abstract tasks as data sizes and CPU cycles and use a single delay or energy reward, mostly in simulation.

\textbf{Video analytics offloading.} DeepDecision~\cite{r13} chooses between a small device model and a large server model based on accuracy, bitrate, battery, and bandwidth, and EdgeDuet~\cite{r25} detects large objects on the device and small objects at the edge, validated over LTE and Wi-Fi. Both rely on optimization or heuristics for a single vision task.

\textbf{Edge-cloud LLM and VLM inference.} Routing between small and large models by cascades (FrugalGPT~\cite{r28}) or learned routers (RouteLLM~\cite{r30}) optimizes API cost and quality but not device energy or the link. Wireless-aware routing weighs query semantics with communication and computation costs~\cite{r32}, controls cost and risk at the edge~\cite{r33}, or routes VQA requests by input complexity (INAR-VL~\cite{r34}). Split inference sends only query-relevant features~\cite{r35} or partitions an LLM across devices~\cite{r37}, but needs the link for every query.

\begin{table}[t]
\centering
\caption{Comparison with prior offloading and routing work. \cmark: addressed, \pmark: partially addressed, \xmark: not addressed.}
\label{tab:related}
\scriptsize
\setlength{\tabcolsep}{1.6pt}
\begin{tabular}{lccccccc}
\toprule
\textbf{Work} & \textbf{Link} & \textbf{Models} & \textbf{Quality} & \textbf{Energy} & \textbf{RRC} & \textbf{Load} & \textbf{5G} \\
\midrule
Kim et al.~\cite{r15} & \cmark & \xmark & \xmark & \cmark & \xmark & \xmark & \xmark \\
DROO~\cite{r14}, \cite{r22} & \cmark & \xmark & \xmark & \pmark & \xmark & \pmark & \xmark \\
PPO~\cite{r24,r55,r56} & \pmark & \pmark & \xmark & \pmark & \xmark & \pmark & \xmark \\
Nieto et al.~\cite{r57} & \pmark & \xmark & \xmark & \cmark & \xmark & \pmark & \cmark \\
DeepDecision~\cite{r13} & \cmark & \pmark & \pmark & \cmark & \xmark & \xmark & \xmark \\
LLM routing~\cite{r28,r30} & \xmark & \cmark & \cmark & \xmark & \xmark & \xmark & \xmark \\
Wireless routing~\cite{r32,r33,r34} & \cmark & \cmark & \cmark & \pmark & \xmark & \xmark & \xmark \\
Split inference~\cite{r35,r37} & \pmark & \xmark & \cmark & \xmark & \xmark & \xmark & \xmark \\
\midrule
\textbf{Ours} & \cmark & \cmark & \pmark & \cmark & \pmark & \cmark & \cmark \\
\bottomrule
\end{tabular}
\\[2pt]
\parbox{\columnwidth}{\scriptsize Quality \pmark: one average score per model and question type. RRC \pmark: inferred from idle time because the modem does not report it. Load \pmark: a device task queue~\cite{r22,r55} or available server resources~\cite{r57}; neither includes the load of other clients on a shared server.}
\end{table}

To our knowledge, no prior work jointly measures, on a real 5G SA link, the RRC state changed by offloading and the shared-server load created by other clients, and validates a policy that observes them (Table~\ref{tab:related}).

\section{Measurement Study on Private 5G MEC}\label{sec:measure}

\subsection{Testbed and Tasks}\label{sec:testbed}\label{sec:tasks}

The device is an NVIDIA Jetson AGX Orin with an HM962 5G modem. To vary its computing capability, we use the three official power modes, MAXN (12 CPU cores at 2.2GHz, 8 GPU TPCs at 1.3GHz), 30W (8 cores at 1.73GHz, 4 TPCs at 612MHz), and 15W (4 cores at 1.11GHz, 3 TPCs at 408MHz), representing high-, mid-, and low-power robots, with clocks fixed at the mode maximum. The network is a private 5G SA network on 100MHz of band n79, and the MEC server (RTX A6000) is wired to the UPF, so the only wireless segment is one radio hop (ping RTT about 9ms). Each request logs the end-to-end latency and the modem's RSRP, RSRQ, SINR, and RSSI. As in~\cite{r15}, device energy is the mean INA3221 power (20Hz) divided by the throughput, summed over four Jetson board rails (VDD\_GPU\_SOC, VDD\_CPU\_CV, VIN\_SYS\_5V0, and the DRAM rail VDDQ\_VDD2\_1V8AO), without the 5G modem; omitting the DRAM rail would raise the remote-to-local energy ratio by 2-5\% and change no decision. The modem, measured separately with a USB power meter (AVHzY CT-3, 10Hz), drew 0.61W during local execution and 1.24W during offloading; with it, a PE frame costs 1,022mJ local and 751mJ remote, and a VLM offload about 0.5J more (4\% of the remote board energy).

We use four tasks whose compute differs widely (Table~\ref{tab:tasks}): object detection (OD, YOLOv7-tiny~\cite{r46}), pose estimation (PE, HRNet-w32~\cite{r47} on each detected person), and two VLM tasks. OD and PE run the same TensorRT FP16 models on both sides. For VLM the remote model is always Qwen2.5-VL-7B (Q8\_0)~\cite{r6} and the local model is Qwen2.5-VL-3B (Q4) for the small VLM (VLM-S) or SmolVLM2-2.2B (Q4)~\cite{r48} for the tiny VLM (VLM-T), all served by llama.cpp~\cite{r49}, so offloading also changes answer quality. VLM questions are short, medium, or long (up to 16, 48, and 128 tokens). Compute $c$ is the GFLOPs per request: 13.8 for OD, 42.2 for PE with four people, and, by the estimate $2Pn$ ($P$ parameters including the vision encoder, $n$ input and output tokens including about 300 visual tokens of a $640\times360$ image), 2,513/2,813/3,465 for short/medium/long VLM-S questions and about 1,800 for VLM-T.

\begin{table}[t]
\centering
\caption{Tasks, compute $c$ (GF: GFLOPs), latency constraint $\delta_q$, and remote-to-local (R/L) ratios on MAXN at RSRP $\approx -74$dBm. R/L $<1$ favors offloading.}
\label{tab:tasks}
\scriptsize
\setlength{\tabcolsep}{2pt}
\begin{tabular}{llrrrrr}
\toprule
\textbf{Task} & \textbf{Local / remote} & \textbf{$c$ (GF)} & \textbf{$\delta_q$} & \textbf{Local (ms)} & \textbf{Lat. R/L} & \textbf{E R/L} \\
\midrule
OD & YOLOv7-tiny & 13.8 & 300ms & 20.1 & 1.88 & 1.20 \\
PE & + HRNet-w32 & 42.2$^\dagger$ & 300ms & 45.6 & 1.25 & 0.74 \\
VLM-T & SmolVLM2 / Qwen-7B & 1800$^\S$ & 1/2/4s & 831 & 0.98 & 0.32 \\
VLM-S short & Qwen-3B / Qwen-7B & 2513 & 1s & 530 & 0.47 & 0.13$^\ddagger$ \\
VLM-S med. & Qwen-3B / Qwen-7B & 2813 & 2s & 1286 & 0.63 & 0.17$^\ddagger$ \\
VLM-S long & Qwen-3B / Qwen-7B & 3465 & 4s & 2762 & 0.65 & 0.17$^\ddagger$ \\
\bottomrule
\end{tabular}
\\[2pt]
\parbox{\columnwidth}{\scriptsize $^\dagger$13.8 + 7.1GFLOPs per person (four people). VLM-S $c$: $2Pn$ ($P$ = 3.75B incl.\ vision encoder; $n$ = prompt with about 300 visual tokens + answer tokens). $^\S$Medium-length estimate; p50 over 77 common questions of all types. $^\ddagger$Power ratio $\times$ latency ratio. 30W/15W: Fig.~\ref{fig:compute}.}
\end{table}

\subsection{Offloading Gain versus Compute}\label{sec:gain}

\begin{figure}[t]
\centering
\includegraphics[width=0.85\columnwidth]{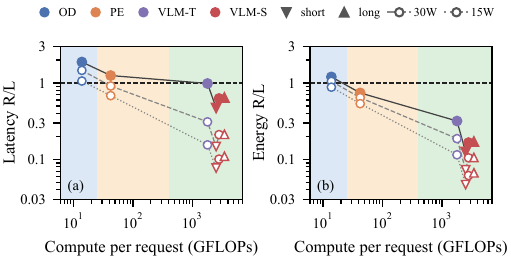}
\caption{Remote-to-local ratio of (a) latency and (b) device energy versus per-request compute under good signal (RSRP $\approx -74$dBm); a ratio below 1 favors offloading. Shading: the three regimes without server load.}
\label{fig:compute}
\end{figure}

Fig.~\ref{fig:compute} shows the remote-to-local latency and energy ratios as a function of compute under good signal conditions, using the means of three runs on the common inputs, which agreed within $\pm$2\%. The ratio falls steeply with compute, and without server load the requests (task $\times$ question type) fall into three regimes: \emph{local-favored}, \emph{state-dependent} (the decision can flip with the link or server state), and \emph{offload-favored}. OD is local-favored on MAXN, PE is state-dependent, and both shift toward offloading on 30W and 15W. VLM-S favors offloading in every mode, yet short questions on MAXN differ by only 280ms (0.53s vs.\ 0.25s) and are state-dependent; on 30W and 15W the local 3B exceeds even the 1s constraint of short questions (1.70s and 3.28s). Locally on MAXN, VLM-T meets the constraints of all three types (0.60s, 0.83s, and 1.31s), which gives a local escape under server congestion; on 30W and 15W even its short questions exceed 1s. Slow execution also costs energy: from MAXN to 15W the local energy per question rises from 77J to 125J.

\subsection{Link Degradation, RRC State, and Server Load}\label{sec:factors}

\begin{figure}[t]
\centering
\includegraphics[width=0.85\columnwidth]{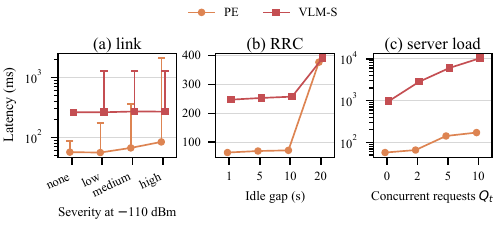}
\caption{Factors beyond compute (MAXN, remote). (a) Latency at RSRP -110dBm under uplink outages of increasing severity (none: -80dBm without outage); p50 with p99 whiskers. (b) Latency of the first offload after an idle gap (inferred RRC transitions). (c) Latency versus concurrent requests from other clients (VLM-S: mean over question types; one 7B server for VLM-S and VLM-T).}
\label{fig:linkload}
\end{figure}

We emulate weak signal by uplink outages (tc filter) that follow a field RSRP trace (-109dBm to -77dBm), more frequent as the signal weakens, at three severities (low, medium, high). Down to -100dBm the latency did not change, and at -105dBm and -110dBm only the tail grew (Fig.~\ref{fig:linkload}(a)). For the state-dependent PE at -110dBm and high severity, the p99 latency reached 2.0-2.5s, 13-16\% of requests violated the 300ms latency constraint, and the offloading energy (1.8-1.9J) exceeded local execution (1.0-1.3J), flipping the decision to local. VLM-S violations also rose to at most 15\%, but the remote energy (21J) stayed far below the local energy (77J), so its decision did not flip. This is a stress test on the real 5G SA link without mobility or handover; the simulator synthesizes SINR from RSRP (Section~\ref{sec:sim}).

Two further factors change the remote cost. First, choices affect later costs: after a run of local executions the modem drops to an RRC power-saving (inactive) state, and the next offload pays a reconnection delay~\cite{r17}. Reconnections were rare for idle gaps up to 10s (1 of 180) but occurred in about 60\% of cases (62 of 105) at 11-20s, adding about 330ms for PE and 300ms for VLM-S (Fig.~\ref{fig:linkload}(b)). The modem does not report its RRC state, so we attribute this delay to an inferred transition to this state (Section~\ref{sec:state}). Second, server load from other clients raises the remote latency: with 0, 2, 5, and 10 concurrent clients the median remote latency grew from 57ms to 172ms for PE and from 0.83s to 10.1s for the shared 7B server (Fig.~\ref{fig:linkload}(c)). The load comes from $Q_t$ closed-loop virtual clients (Section~\ref{sec:system}), each sending its next request on a response; the server runs one request at a time and queues the rest FCFS. Unlike link degradation, server load flips even VLM-S.

\subsection{Observations and Design Requirements}\label{sec:obs}

\begin{description}
\item[O1.] \textbf{Compute is the primary factor.} With a short, stable 5G MEC path, the decision depends mainly on how heavy the model is, and link quality changes it only in the state-dependent regime (O3). $\rightarrow$ Tasks are represented by their compute $c$, so one policy handles all tasks.
\item[O2.] \textbf{The device power mode changes the decision for state-dependent tasks,} and latency and energy can disagree. $\rightarrow$ The power mode is part of the state, and the reward combines energy with the latency constraint.
\item[O3.] \textbf{Link degradation flips only the state-dependent PE, whereas server load flips even VLM-S; the RRC state depends on earlier choices.} $\rightarrow$ The RRC state and the server load join the link metrics in the state, and decisions are made sequentially.
\item[O4.] \textbf{Question types differ in latency constraint and quality.} The remote 7B beats the local 3B for every type (Section~\ref{sec:load}); SmolVLM2 is close to 3B for short and medium questions and lower for long ones. $\rightarrow$ The reward uses per-type latency constraints $\delta_q$ and a quality term, while the question type itself is not a policy input.
\end{description}

\section{Problem Formulation}\label{sec:formulation}

\subsection{System Model}\label{sec:system}

The robot processes inference requests one at a time, indexed by $t = 1, 2, \dots$ with arrival time $a_t$: a camera frame for OD and PE, or a question for the VLM family. Request $t$ has a task $k_t \in \mathcal{K} = \{\mathrm{OD}, \mathrm{PE}, \mathrm{VLM\text{-}T}, \mathrm{VLM\text{-}S}\}$ and the compute $c_t$ (GFLOPs) of the model that runs it, a per-task constant that, for the VLM family, is the mean over the three question types. A VLM-family request also has a question type $q_t \in \mathcal{Q} = \{\mathrm{short}, \mathrm{medium}, \mathrm{long}\}$. On arrival the device observes
\begin{equation}
s_t = \big(\ell_t,\ c_t,\ p,\ \rho_t,\ Q_t,\ x_{t-1},\ \tau_t\big),
\label{eq:state}
\end{equation}
where $\ell_t = (\mathrm{RSRP}_t, \mathrm{RSRQ}_t, \mathrm{SINR}_t, \mathrm{RSSI}_t)$ are the link metrics, $p \in \{\mathrm{MAXN}, \mathrm{30W}, \mathrm{15W}\}$ the power mode, $\rho_t \in \{0, 1\}$ the RRC state (1: connected), $Q_t$ the number of concurrent requests from other clients at the MEC server, $x_{t-1}$ the previous action, and $\tau_t$ the time since the last offload. It then chooses $x_t \in \{0, 1\}$ (0: local, 1: offload). The question type $q_t$ enters only the latency constraint and the quality, not the state. Since $k_t$ is not in the state either, $c_t$ alone distinguishes the tasks.

The components of $s_t$ evolve in two ways. The link metrics and the load are \emph{exogenous}: $\ell_t$ follows the field RSRP trajectory (Section~\ref{sec:factors}) and $Q_t = Q(a_t)$ follows the load that other clients place on the server; the robot's own request is not counted in $Q_t$, in the simulator and in the real-device experiment (Section~\ref{sec:real}). The RRC state is \emph{endogenous}: with $\Delta_t = a_{t+1} - a_t$,
\begin{equation}
\tau_{t+1} = (1 - x_t)(\tau_t + \Delta_t), \qquad \rho_{t} = \mathbf{1}[\tau_t < T_t],
\label{eq:rrc}
\end{equation}
where $T_t$ is the inactivity timer set by the network and unknown to the device. Since reconnections were rare for idle gaps up to 10s (1 of 180, Section~\ref{sec:factors}), the analysis uses only the lower bound $T_t \ge 10.5$s (the simulator's timer is given in Section~\ref{sec:sim}). The simulator measures $\tau$ from the request after an offload and the device from the offload response, so simulated idle periods are shorter by one question gap (about 3s; 67ms for frames), a difference well below the 10.5s bound. An offload with $\rho_t = 0$ pays a reconnection delay $\Delta_{\mathrm{RRC}}$ (about 330ms for gRPC and 300ms for HTTP), and one with $\rho_t = 1$ does not. Actions also shift later arrival times, since the next frame or question waits for the current result, and hence the load phase that later requests meet. Thus $\rho$ and $\tau$ depend on the actions and affect later costs, whereas $Q_t$ can only be observed.

\subsection{Cost and Quality Functions}\label{sec:cost}

The outcome of request $t$ under $x_t$ has three parts. The Jetson board energy $e_t(x_t)$ (four INA3221 rails, 5G modem excluded, Section~\ref{sec:testbed}) is nearly constant for local execution of a given task and question type, whereas for offloading it includes transmission and waiting and depends on $\ell_t$, $\rho_t$, and $Q_t$. Since energy differs by two orders of magnitude across tasks (about 0.4J for OD and 77J for VLM-S on MAXN) and fivefold across VLM question types, we normalize it as $\tilde{e}_t = e_t(x_t)/\bar{e}_L(k_t, q_t, p)$, where $\bar{e}_L$ is the mean local energy for that task, type, and power mode. The delay $d_t(x_t)$ has a latency constraint $\delta_q$: 300ms for OD and PE~\cite{r15}, and 1s, 2s, and 4s for short, medium, and long questions. The quality $\hat{g}(m, q) \in [0, 1]$ is the pre-graded mean score of the execution model $m$ for type $q$ (Section~\ref{sec:setup}): local 3B 0.678/0.40/0.33, local SmolVLM2 0.687/0.40/0.29, and remote 7B 0.703/0.50/0.43 for short/medium/long questions; OD and PE use the same model on both sides and have no quality term.

\subsection{Objective Function}\label{sec:objective}

A per-request hard constraint $d_t \le \delta_{q_t}$ cannot be guaranteed when the link and the load change unpredictably, so we pose a constrained MDP that minimizes the mean per-request cost $J(\pi)$ of a policy $\pi$ (a rule that maps the state to an action) subject to a bound on the long-run violation rate $V(\pi)$:
\begin{equation}
\begin{aligned}
\min_{\pi}\ & J(\pi) = \lim_{N\to\infty}\tfrac{1}{N}\textstyle\sum_{t=1}^{N}\mathbb{E}_\pi\big[\tilde{e}_t - \lambda\,\hat{g}(m_t, q_t)\big] \\
\text{s.t.}\ & V(\pi) = \lim_{N\to\infty}\tfrac{1}{N}\textstyle\sum_{t=1}^{N}\Pr_\pi\big[d_t > \delta_{q_t}\big] \le \varepsilon .
\end{aligned}
\label{eq:objective}
\end{equation}
Its Lagrangian relaxation with multiplier $\omega \ge 0$ gives the per-request reward
\begin{equation}
r_t = -\tilde{e}_t - \omega\,\mathbf{1}\big[d_t(x_t) > \delta_{q_t}\big] + \lambda\,\hat{g}(m_t, q_t),
\label{eq:reward}
\end{equation}
and the policy maximizes its value $\mathbb{E}_\pi[\sum_t \gamma^t r_t]$ with discount $\gamma$, whose effective horizon is about $1/(1-\gamma)$ requests. When the long-run averages exist, the value multiplied by $(1-\gamma)$ approaches $-(J + \omega V)$ as $\gamma \to 1$. Since the simulator uses finite measured samples and bounded added delays, the reward is bounded, $-(\tilde{e}_{\max} + \omega) \le r_t \le \lambda$ with $\tilde{e}_{\max}$ the maximum normalized energy, and all values are finite for $\gamma < 1$.

\noindent\textbf{Lemma 1 (role of $\omega$).} Let $\pi_\omega$ minimize $J(\pi) + \omega V(\pi)$ among the stationary policies for which the long-run averages $J$ and $V$ exist. (i) $\pi_\omega$ solves (\ref{eq:objective}) with $\varepsilon = V(\pi_\omega)$. (ii) For $\omega_1 < \omega_2$, $V(\pi_{\omega_2}) \le V(\pi_{\omega_1})$ and $J(\pi_{\omega_2}) \ge J(\pi_{\omega_1})$.

\noindent\textit{Proof.} (i) If some $\pi$ had $V(\pi) \le V(\pi_\omega)$ and $J(\pi) < J(\pi_\omega)$, then $J(\pi) + \omega V(\pi) < J(\pi_\omega) + \omega V(\pi_\omega)$, a contradiction. (ii) Optimality of each policy gives $J_1 + \omega_1 V_1 \le J_2 + \omega_1 V_2$ and $J_2 + \omega_2 V_2 \le J_1 + \omega_2 V_1$; adding them yields $(\omega_2 - \omega_1)(V_2 - V_1) \le 0$, and substituting back gives $J_2 \ge J_1$.

Thus $\omega$ selects a point on the energy-violation trade-off curve, and $\varepsilon$ is the allowed violation rate. The baseline~\cite{r15} approximates $\varepsilon = 0$ with a fixed $\omega$; we likewise fix $\omega$, so by Lemma~1(i) $\varepsilon = V(\pi_\omega)$. PPO only approximates $\pi_\omega$, so Lemma~1 interprets $\omega$ and does not guarantee a given $\varepsilon$; the violation rates in Section~\ref{sec:eval} are measured on the evaluation trajectories. For a required $\varepsilon$, Lemma~1(ii) allows finding the smallest $\omega$ that meets it. We adopt $\omega = 10$ from~\cite{r15}; retraining with $\omega \in \{1, 10, 100\}$ shows the violation rate falling and $J$ and the energy rising monotonically, with $\omega = 10$ at the knee (Section~\ref{sec:load}). The single $\lambda = 0.567$ (the mean of the per-type values 0.2, 0.5, and 1.0) was checked over 0-2 times its value.

\subsection{Relation to the Baseline and Sequential Structure}\label{sec:relation}

With $\lambda = 0$ and a single task, $\bar{e}_L$ is constant and (\ref{eq:reward}) becomes $r_t = -\frac{1}{\bar{e}_L}\big(e_t + \omega\bar{e}_L\,\mathbf{1}[d_t > \delta]\big)$, the per-request cost of the baseline with penalty $\omega' = \omega\bar{e}_L$; Section~\ref{sec:eval} therefore compares both on the same objective. By (\ref{eq:rrc}), however, the action changes the next RRC state, so the problem is a sequential decision problem rather than a bandit. The RRC part is Markov in $(\rho_t, \tau_t)$, but unobserved exogenous state such as the remaining length of a load phase remains, so we treat the whole as an approximate MDP $(\mathcal{S}, \{0,1\}, \mathcal{P}, r, \gamma)$. With $\gamma = 0$ it reduces to a contextual bandit, which LinUCB~\cite{r50} solves. Proposition~1 shows what this reduction loses.

\noindent\textbf{Proposition 1 (myopic lock-in).} Consider a stationary stream of one task under good signal and no load, with mean rewards $r_L$ for local execution and $r_R > r_L$ for an offload in the connected state, inter-arrival times below the minimum timer ($\Delta_t < 10.5$s), and a reconnection that makes the first offload after idling violate its latency constraint ($d_R + \Delta_{\mathrm{RRC}} > \delta$) while connected offloads do not. Let $\tilde{e}_{\mathrm{RRC}}$ be the additional normalized energy the board uses during the reconnection. Starting in the power-saving state, the myopic policy ($\gamma = 0$) executes locally forever whenever $r_R - r_L < \omega + \tilde{e}_{\mathrm{RRC}}$, whereas for the discounted objective offloading from the first request is strictly better than always-local, and thus optimal, if and only if
\begin{equation}
\frac{r_R - r_L}{1 - \gamma} > \omega + \tilde{e}_{\mathrm{RRC}} .
\label{eq:lockin}
\end{equation}

\noindent\textit{Proof.} In the power-saving state the immediate rewards are $r_L$ for local execution and $r_R - \omega - \tilde{e}_{\mathrm{RRC}}$ for offloading, so the myopic policy executes locally when $r_R - r_L < \omega + \tilde{e}_{\mathrm{RRC}}$; by (\ref{eq:rrc}) $\tau$ keeps growing and the RRC state stays power-saving, so the choice repeats forever. For the discounted objective ($\gamma > 0$), always-local has the value $r_L/(1-\gamma)$. Offloading from the first request pays the reconnection penalty and energy only on that request; afterwards the gaps stay below the timer, so $\rho = 1$ holds and every request earns $r_R$, giving the value $(r_R - \omega - \tilde{e}_{\mathrm{RRC}}) + \gamma r_R/(1-\gamma) = r_R/(1-\gamma) - \omega - \tilde{e}_{\mathrm{RRC}}$. The value of offloading after $i$ local executions is the average of these two values weighted by $1-\gamma^i$ and $\gamma^i$, so it cannot be optimal. Hence offloading from the first request is strictly better than always-local if and only if $r_R/(1-\gamma) - \omega - \tilde{e}_{\mathrm{RRC}} > r_L/(1-\gamma)$, i.e., (\ref{eq:lockin}); with equality the two values are equal.

For PE on MAXN, offloading saves energy ($r_R - r_L \approx 1 - 0.74 = 0.26$), the remote delay of about 57ms plus $\Delta_{\mathrm{RRC}} \approx 330$ms exceeds 300ms, and the camera frame interval is 67ms (15fps; frames that arrive during processing are skipped), so the conditions hold; the additional energy over the 330ms reconnection is $\tilde{e}_{\mathrm{RRC}} \approx 4.1$. Since $0.26 < \omega + \tilde{e}_{\mathrm{RRC}} \approx 14.1$, the greedy myopic policy that does not explore locks into local execution, while (\ref{eq:lockin}) holds for $1/(1-\gamma) > (\omega + \tilde{e}_{\mathrm{RRC}})/(r_R - r_L) \approx 54$ requests. Our $\gamma = 0.995$ (an effective horizon of 200 requests) satisfies it; LinUCB with $\alpha = 1$ shows this lock-in in S1 and S2, where one penalized offload stopped further exploration (Section~\ref{sec:ablation}).

\section{Congestion-Aware Sequential Offloading}\label{sec:method}

\subsection{Overview}\label{sec:overview}

\begin{figure}[t]
\centering
\includegraphics[width=0.85\columnwidth]{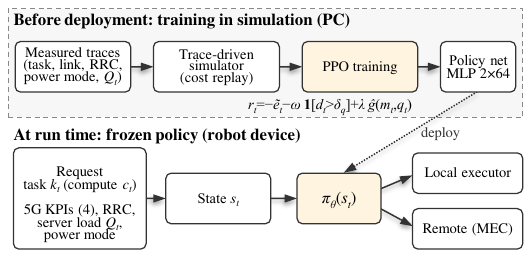}
\caption{Congestion-aware sequential offloading. A PPO policy is trained in a trace-driven simulator where offloading changes the RRC state of later requests; at run time, the frozen policy network on the device selects local or remote execution from the observed state.}
\label{fig:policy}
\end{figure}

Fig.~\ref{fig:policy} shows the two phases. Before deployment, a policy is trained with PPO~\cite{r18} in a simulator built from the traces of Section~\ref{sec:measure}; after deployment, a small frozen policy network on the device maps the state (\ref{eq:state}), formed from the modem's 5G KPIs, the RRC estimate, the server load, the request's compute, and the power mode, to local execution or offloading. Unlike the single-task baseline~\cite{r15}, one policy serves models whose compute differs by more than two orders of magnitude (OD, PE, VLM-T, and VLM-S): tasks enter through $c_t$, the RRC state $(\rho_t, \tau_t)$ and the load $Q_t$ are task-independent state, and the per-task normalization puts the rewards (\ref{eq:reward}) on one scale.

\subsection{State Representation}\label{sec:state}

The state (\ref{eq:state}) gives 12 inputs: the four link metrics scaled min-max over $[-120, -70]$dBm (RSRP), $[-20, -8]$dB (RSRQ), $[-10, 32]$dB (SINR), and $[-110, -60]$dBm (RSSI); the compute as $(\log_{10} c_t - 1)/3$ with $c_t$ = 13.8, 42.2, about 1{,}875 (mean over question types), and 2{,}930GFLOPs for OD, PE, VLM-T, and VLM-S; the power mode one-hot (3); $\rho_t$ and $x_{t-1}$ as is; $Q_t/10$; and $\min(\tau_t, 20\mathrm{s})/20\mathrm{s}$. Since $c_t$ of the VLM family is averaged over the question types, the type does not leak into the policy; adding it as an input did not improve the results (S4 reward $-2.598$ vs.\ $-2.538$).

The elapsed time $\tau_t$ makes the endogenous part Markov. By (\ref{eq:rrc}), given $\tau_t$ the RRC state is power-saving with probability $\Pr[T_t \le \tau_t] = \min\{1, \max\{0, (\tau_t - 10.5)/9.5\}\}$, and $\tau_{t+1}$ depends on $(\tau_t, x_t, \Delta_t)$, so $(\rho_t, \tau_t)$ summarizes all that past actions contribute to future costs. In the simulator, $\rho_t$ is the true state set by the timer drawn for each idle period (Section~\ref{sec:sim}); on the real device (Section~\ref{sec:real}), the modem does not report $\rho_t$ and bandwidth-based inference lags the release, so $\rho_t$ is estimated as $\mathbf{1}[\tau_t < 10.5\mathrm{s}]$. In all experiments $Q_t$ is read from the load schedule; delayed observations and updates carried only by offload responses are discussed in Section~\ref{sec:limits}.

\subsection{Simulator and Training}\label{sec:sim}

The simulator samples the base energy and latency of the chosen action from empirical distributions of about 100{,}000 measured requests (Section~\ref{sec:gain}) grouped by task, question type, power mode, and location; VLM-S (local Qwen2.5-VL-3B) and VLM-T (local SmolVLM2-2.2B) share the cost, queueing, and quality of the same remote Qwen2.5-VL-7B server. Along a field RSRP trajectory, an offload adds an upload-time increase and a latency drawn from the measured outage-injection distribution for the RSRP range and severity (Section~\ref{sec:factors}). The upload rate follows the attenuated Shannon model~\cite{r51}, $B = B_0\,\eta(\mathrm{SINR})/\eta_{\max}$ with $\eta = \min\{\eta_{\max}, 0.4\log_2(1 + \mathrm{SINR})\}$ and $\eta_{\max} = 2.0$bps/Hz ($\eta = 0$ below $-10$dB), where $\mathrm{SINR} = \min(32, \mathrm{RSRP} + 113)$dB and $B_0 = 355$Mbps is the uplink rate measured under good signal. The RRC state follows (\ref{eq:rrc}) with the timer redrawn from $\mathcal{U}(10.5, 20)$s for each idle period. The reconnection delay is the measured one, and the load adds the measured waiting time for $Q_t$ scaled by $\mathcal{U}(0.5, 1.5)$. The additional offloading energy is the mean remote-execution power times the additional latency. Training trajectories are synthesized from 30-90s segments cut from the same three field drives used for evaluation, which are reversed, shuffled, and shifted by $-10$ to $+6$dB. The original trajectory (648s) and its $-8$dB cell-edge version were not used for training in their original order, but they come from the same drives, so the evaluation is a replay check on the same site (Section~\ref{sec:limits}). Training streams mix OD, PE, and the three VLM question types (main model), and a policy also trained with VLM-T (PPO-T) is the reference for generalization to VLM-T (Section~\ref{sec:ablation}).

PPO maximizes the clipped surrogate objective of~\cite{r18} with the reward (\ref{eq:reward}):
\begin{equation}
\begin{aligned}
\mathcal{L}(\theta, \phi) = {}& \hat{\mathbb{E}}_t\big[\min\big(u_t\hat{A}_t,\ \mathrm{clip}(u_t, 1{-}\epsilon_{\mathrm{clip}}, 1{+}\epsilon_{\mathrm{clip}})\hat{A}_t\big)\big] \\
& - c_v\,\hat{\mathbb{E}}_t\big[(v_\phi(s_t) - \hat{R}_t)^2\big] + c_h\,\hat{\mathbb{E}}_t\big[\mathcal{H}(\pi_\theta(\cdot \mid s_t))\big],
\end{aligned}
\label{eq:ppo}
\end{equation}
where $\pi_\theta$ and $v_\phi$ are the policy and value networks, $u_t = \pi_\theta(x_t \mid s_t)/\pi_{\theta_{\mathrm{old}}}(x_t \mid s_t)$ is the probability ratio to the pre-update policy, clipped by $\epsilon_{\mathrm{clip}}$ to limit each update, $\hat{A}_t = \sum_{j \ge 0} (\gamma\lambda_{\mathrm{GAE}})^j \zeta_{t+j}$ is the GAE advantage with temporal-difference error $\zeta_t = r_t + \gamma v_\phi(s_{t+1}) - v_\phi(s_t)$, $\hat{R}_t = \hat{A}_t + v_\phi(s_t)$ is the value target, the entropy $\mathcal{H}$ keeps the policy exploring, and $c_v$ and $c_h$ weight the last two terms. We use $\gamma = 0.995$, $\epsilon_{\mathrm{clip}} = 0.2$, $c_v = 0.5$, $c_h = 0.01$, $\lambda_{\mathrm{GAE}} = 0.95$, rewards scaled by 0.1, rollouts of 8,192 requests, 10 epochs with minibatches of 256, learning rate $3\times10^{-4}$, gradient norm clipped at 0.5, and $8.2\times10^{6}$ requests in total; the policy and value networks are two-layer MLPs with 64 units (5{,}122 policy parameters), trained with five seeds. We chose only $\gamma$ (Section~\ref{sec:relation}); the other hyperparameters are common PPO defaults~\cite{r18}.

\subsection{Deployment}\label{sec:deploy}

Only the policy network is deployed, with the most probable action selected. A decision takes 0.026ms on the device CPU in the slowest 15W mode (SW-UCB 0.018ms), negligible against processing times of tens of milliseconds to seconds, and the deployed network reproduces the simulator's decisions on all evaluated S4 states. There is no online training, so no exploration cost, but the policy may be suboptimal where conditions differ from the training traces (Section~\ref{sec:discussion}).

\subsection{Comparison Variant: LinUCB}\label{sec:linucb}

The baseline SW-UCB~\cite{r15} and the proposed policy differ in two ways, the state and sequential decisions, so to separate their effects we add LinUCB~\cite{r50}, which uses the same 12 inputs. LinUCB learns a linear reward model per action online with upper-confidence exploration and ignores the action's effect on the next state ($\gamma = 0$ in Section~\ref{sec:relation}): SW-UCB vs.\ LinUCB isolates the state, and LinUCB vs.\ PPO shows the combined effect of sequential decisions and pretraining.

\section{Evaluation}\label{sec:eval}

\subsection{Experimental Setup}\label{sec:setup}

\textbf{Data and answer quality.} We use 1,214 keyframes from 20 sequences of the five most crowded venues in the validation split of the ``Robot-Perspective Driving Video (Advanced)'' dataset (NIA, AI-Hub); OD and PE use the crowded part of the video because PE scales with the number of people. For each keyframe, Claude Opus 5.5 generates a short, a medium, and a long question with reference answers. The prompt asks for questions with one clearly correct answer from the 640$\times$360 robot-camera frame, with no identity attributes and an answerability flag: short ones on the venue type, the number of people, or an obstacle; medium ones on the layout or a person (within 35 words); long ones on whether and why the robot can move on safely (within 90 words). After dropping unanswerable frames and short questions the generator flagged as ambiguous, the 397 venue answers come from the dataset labels; for the other short answers, a human checked 50 random ones (92\% correct as generated) and judged the 92 that disagreed with another vendor's model, fixing 27 (estimated accuracy at least 98\%). This leaves 3,318 questions (902 short, 1,208 medium, 1,208 long). Short questions are scored by exact match; medium and long ones are scored 1-5 by an LLM judge~\cite{r41} from another vendor (gpt-6.1-sol), since judges favor their own model's text~\cite{r43}, and by a second judge (Claude Sonnet 5.5). The mean of the two scores, mapped to $[0,1]$, gives $\hat{g}(m,q)$ (Section~\ref{sec:cost}). The second judge shares the generator's vendor but scored lower, and GPT alone gives the same ranking (Section~\ref{sec:load}); the judges agree with Spearman correlation 0.789 and weighted $\kappa = 0.714$.

\textbf{Compared policies.} Always-Local and Always-Remote; the field rule (thresholds on 5G indicators from a field trial); SW-UCB of the reference study~\cite{r15} with RSSI (original) or RSRP as state, and with per-type latency constraints $\delta_q$; LinUCB (Section~\ref{sec:linucb}; $\alpha = 20$, with $\alpha = 1$ from the original grid); and a Myopic Oracle. SW-UCB uses $\omega' = \omega\bar{e}_L$ (Section~\ref{sec:relation}) and, among the reference-study candidates, windows $W \in \{1{,}000, 2{,}000\}$ requests and exploration weights $\xi \in \{0.5, 2\}$, the values chosen on the training trajectories (RSSI $W = 2{,}000$, RSRP $W = 1{,}000$, $\xi = 0.5$). The real-device SW-UCB uses the same setting and updates with a per-request energy of reference power times measured latency, which avoids the power-sensor delay. The Myopic Oracle is a request-level hindsight reference that knows the outcomes of both actions and picks the one with the larger reward~(\ref{eq:reward}), with RRC assumed active, since otherwise it would see only the penalty and reconnection energy of the first offload after power saving and lock into local execution as in Proposition~1. It ignores how the current action affects later requests, so a sequential policy can beat it in S4 (Section~\ref{sec:s3s4}).

\textbf{Scenarios and metrics.} In the simulator of Section~\ref{sec:sim}, we replay in order the original field drives, which training used only as cut and synthesized segments: S1, the reference-study condition (a single PE stream, only the link varies); S2, task switching OD$\rightarrow$PE$\rightarrow$VLM-S; S3, link degradation with a mix of the three question types; and S4, S3 plus server-load phases ($Q_t$ = 0/2/5/10). Each scenario runs on the field RSRP trajectory (648s) and on a cell-edge version 8dB lower, with four seeds, three power modes, and three outage severities. Metrics are the per-request reward, normalized energy, violation rate, and answer quality; the reward uses the per-type weights $\lambda_q = 0.2/0.5/1.0$, whose mean is the training value 0.567; rescoring with 0.567 raises all S3 and S4 rewards by about 0.035 and leaves every ranking unchanged.

\begin{table}[t]
\centering
\caption{Policy comparison: energy normalized to Always-Local (E) / latency violation rate in \% (V), and S3 long-question quality ($\hat{g}_{\mathrm{long}}$); mean over 3 power modes $\times$ 3 severities $\times$ 4 seeds.}
\label{tab:main}
\scriptsize
\setlength{\tabcolsep}{1.7pt}
\begin{tabular}{lcccccc}
\toprule
\textbf{Policy} & \textbf{$\ell$/$\rho$/$Q$} & \textbf{S1 E / V} & \textbf{S2 E / V} & \textbf{S3 E / V} & \textbf{S4 E / V} & \textbf{S3 $\hat{g}_{\mathrm{long}}$} \\
\midrule
Always-Local & -- & 1.00 / \textbf{0.0} & 1.00 / 0.6 & 1.00 / 66.7 & 1.00 / 66.7 & 0.330 \\
Always-Remote & -- & \textbf{0.65} / 0.1 & 0.87 / \textbf{0.1} & \textbf{0.10} / \textbf{0.0} & \textbf{0.43} / 37.2 & \textbf{0.430} \\
Field rule & \cmark\xmark\xmark & 0.81 / 0.1 & 0.94 / 0.2 & 0.47 / 24.2 & 0.72 / 53.9 & 0.390 \\
SW-UCB (RSSI)~\cite{r15} & \cmark\xmark\xmark & \textbf{0.65} / 0.1 & 0.87 / \textbf{0.1} & 0.14 / 3.1 & 0.48 / 40.4 & 0.426 \\
SW-UCB (RSSI, $\delta_q$) & \cmark\xmark\xmark & \textbf{0.65} / 0.1 & 0.87 / \textbf{0.1} & 0.14 / 3.1 & 0.50 / 39.2 & 0.426 \\
SW-UCB (RSRP) & \cmark\xmark\xmark & 0.66 / 0.1 & 0.87 / \textbf{0.1} & 0.14 / 3.1 & 0.48 / 40.5 & 0.426 \\
LinUCB ($\alpha{=}20$) & \cmark\cmark\cmark & 0.71 / \textbf{0.0} & \textbf{0.86} / \textbf{0.1} & 0.21 / 1.3 & 0.50 / 28.3 & 0.417 \\
LinUCB ($\alpha{=}1$) & \cmark\cmark\cmark & 1.00 / \textbf{0.0} & 1.00 / 0.6 & 0.11 / 0.3 & 0.49 / 27.8 & 0.429 \\
\midrule
\rowcolor{gray!18}\textbf{PPO (ours)} & \cmark\cmark\cmark & 0.72 / 0.1 & 0.88 / \textbf{0.1} & \textbf{0.10} / \textbf{0.0} & 0.44 / \textbf{23.6} & \textbf{0.430} \\
Myopic Oracle & -- & 0.62 / 0.0 & 0.79 / 0.0 & 0.10 / 0.0 & 0.39 / 27.8 & 0.430 \\
\bottomrule
\end{tabular}
\\[2pt]
\parbox{\columnwidth}{\scriptsize $\ell$/$\rho$/$Q$: observes link metrics $\ell_t$ / RRC state $\rho_t$ / server load $Q_t$. LinUCB $\alpha{=}1$: best of $\{0.1, 0.5, 1, 2\}$; $\alpha{=}20$: best of the grid widened to $\{5, 20\}$. Bold: best among implementable policies; shaded: proposed. Myopic Oracle: request-level hindsight reference with RRC assumed active, on its own sample path. Field RSRP trace; the $-8$dB version is in the text. Real-device results: Table~\ref{tab:real}.}
\end{table}

\subsection{Reference Condition and Task Switching (S1, S2)}\label{sec:s1}\label{sec:s2}

\begin{figure*}[t]
\centering
\includegraphics[width=0.85\textwidth]{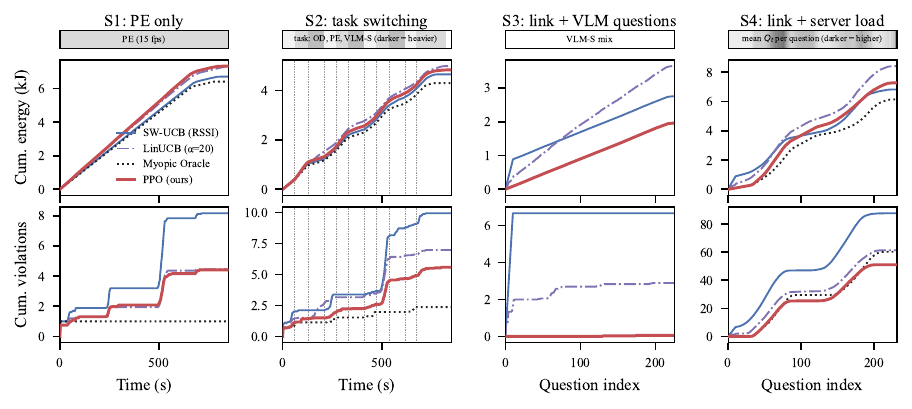}
\caption{Cumulative device energy (top) and latency violations (bottom) in S1-S4, mean over 36 episodes (3 power modes $\times$ 3 severities $\times$ 4 seeds, field trace). S1 and S2 are plotted over time, and S3 and S4 over question index. Strip: task (S2) or mean $Q_t$ per question (S4; darker = higher).}
\label{fig:timeseries}
\end{figure*}

In S1, with a single task and no quality term, the proposed policy and SW-UCB share the same objective (Section~\ref{sec:relation}) and there is little to decide; the proposed policy has a slightly lower reward than Always-Remote and SW-UCB ($-0.725$ vs.\ $-0.655$, $-0.664$; Table~\ref{tab:main}), because it occasionally runs requests locally in anticipation of RRC power saving. In S2 (Fig.~\ref{fig:timeseries}), although SW-UCB pools the statistics of all tasks in one signal bin, its violation rate remained below 0.2\% and field-trajectory rewards were nearly equal: $-0.886$, $-0.879$, $-0.880$ for the proposed policy, SW-UCB, and Always-Remote. On the cell-edge trajectory the proposed policy led: $-0.946$ vs.\ $-1.017$ and $-1.091$.

\subsection{Link Degradation and Server Load (S3, S4)}\label{sec:s3s4}

In S3 the remote 7B beats the local 3B in energy and quality for every type, and link degradation adds some violations (up to 15\%, Section~\ref{sec:factors}) without making local execution better, so offloading is almost always optimal (Table~\ref{tab:main}).

S4 is the core of the evaluation. Neither Always-Remote nor SW-UCB observes the load; SW-UCB learns the remote cost per RSSI bin. With about 32 questions per 130-s load phase over about 18 bins (fewer than two per bin), no window lets this estimate follow the load: windows of 20-200 requests still offloaded 78-100\% of questions under load (41.5-42.7\% violations), whereas a 15-fps PE stream (about 40 frames per bin and phase) does follow it. Both thus keep offloading VLM questions into the queue. The proposed policy, which observes the RRC state and the server load, runs locally under load and cuts the violation rate from 40.4\% (SW-UCB) and 37.2\% (Always-Remote) to 23.6\% with the highest reward ($-2.538$; LinUCB $-3.074$, SW-UCB $-4.251$) at a normalized energy of 0.44 (SW-UCB 0.48). Running locally on the 3B lowered its S4 long-question quality to 0.405 (SW-UCB 0.424). A load-threshold rule that picks a threshold on $Q_t$ per task and power mode on the same training trajectories (e.g., VLM-S runs locally when $Q_t \ge 2$ on MAXN) reached 23.4\% and $-2.516$ in S4, close to the proposed policy (23.6\%, $-2.538$). The gain over SW-UCB thus comes from the observed state rather than from learning itself; the rule, however, needs a threshold for each task and power mode, and at load levels absent from training the proposed policy had fewer violations (Section~\ref{sec:limits}). With only SW-UCB's link inputs (Section~\ref{sec:ablation} A2, Table~\ref{tab:ablation}), the proposed policy still reached 24.5\% and $-2.741$ in S4, but by learning a fixed per-mode rule from the training load (local on MAXN, remote on 30W and 15W) at energy 0.55, so following the load requires $Q_t$. Across seeds the S4 violation rate was 22.4-24.7\%, except one failed seed (33.6\%, reward $-3.52$), excluded from Tables~\ref{tab:main} and~\ref{tab:ablation}.

Per mode, Always-Remote was marginally ahead on 30W, where the local 3B violates even the short-question latency constraint and local execution offers no escape. With VLM-T questions only, which have a local escape on MAXN (PPO-T, trained with VLM-T), the proposed policy ran locally under load on MAXN and reached 0\% violations (Always-Remote 38.2\%, SW-UCB 33.6\%).

\subsection{Server Load, Question Types, and Sensitivity}\label{sec:load}

By load and question type in S4, SW-UCB offloads more than 96\% of questions under load, so 81-99\% of short and medium questions and 26\% of long ones violate at $Q_t = 2$, and at least 98\% at $Q_t \ge 5$. The proposed policy lowers offloading to 31-54\% and violations to 22-66\%; as it does not observe the question type, its offload ratios are nearly equal across types, and the spread of violations comes from the latency constraints.

Scaling $\lambda$ by 0-2 left the results nearly unchanged, since the remote 7B leads in both energy and quality. Retraining with $\omega \in \{1, 10, 100\}$ gave S4 violation rates of 36.3\%, 23.5\%, and 24.8\% at normalized energies of 0.42, 0.44, and 0.52, so $\omega = 10$ sits at the knee of the trade-off, and for every $\omega$ the proposed policy stayed below SW-UCB (40.3-40.9\%) and LinUCB (27.1-39.9\%).

On the evaluation set (Section~\ref{sec:setup}), the local 3B trailed the remote 7B by 2.5pp in short-question accuracy and by about 0.10 in medium and long judge scores, and gave a wrong safety verdict (the judge's lowest safety score, given to every unsafe path called safe) in 9-11\% of long questions (7B: about 5\%). Each judge alone gave the same ranking (3B/SmolVLM2/7B); Claude scored 0.45 and 0.28 points lower than GPT on medium and long questions (1-5 scale).

\subsection{Ablation}\label{sec:ablation}

\begin{table*}[t]
\centering
\caption{Ablation by removed component (S2: task switching, S3: link, S4: load). E: energy normalized to Always-Local; V: latency violation rate (\%); $R$: mean reward (local execution without violations scores about $-1$; each violation costs $\omega = 10$).}
\label{tab:ablation}
\scriptsize
\setlength{\tabcolsep}{3pt}
\newcommand{\w}{\cellcolor{red!15}}
\begin{tabular}{l ccc ccc ccc ccc ccc ccc}
\toprule
 & \multicolumn{6}{c}{\textbf{S2}} & \multicolumn{6}{c}{\textbf{S3}} & \multicolumn{6}{c}{\textbf{S4}} \\
\cmidrule(lr){2-7}\cmidrule(lr){8-13}\cmidrule(lr){14-19}
 & \multicolumn{3}{c}{0dB} & \multicolumn{3}{c}{$-8$dB} & \multicolumn{3}{c}{0dB} & \multicolumn{3}{c}{$-8$dB} & \multicolumn{3}{c}{0dB} & \multicolumn{3}{c}{$-8$dB} \\
\cmidrule(lr){2-4}\cmidrule(lr){5-7}\cmidrule(lr){8-10}\cmidrule(lr){11-13}\cmidrule(lr){14-16}\cmidrule(lr){17-19}
\textbf{Variant} & E & V & $R$ & E & V & $R$ & E & V & $R$ & E & V & $R$ & E & V & $R$ & E & V & $R$ \\
\midrule
\rowcolor{gray!18}\textbf{PPO (full)} & 0.88 & 0.1 & $-0.886$ & 0.93 & 0.2 & $-0.946$ & 0.10 & 0.0 & 0.169 & 0.11 & 0.5 & 0.116 & 0.44 & 23.6 & $-2.538$ & 0.45 & 23.7 & $-2.565$ \\
\midrule
A1: w/o task compute $c$ & 0.87 & 0.1 & $-0.878$ & \w 0.99 & \w 0.6 & \w $-1.043$ & 0.11 & 0.1 & 0.158 & \w 0.16 & \w 0.7 & \w 0.045 & 0.44 & 23.7 & $-2.547$ & \w 0.48 & \w 24.3 & \w $-2.654$ \\
\midrule
A2: w/o $\rho_t$, $\tau_t$, $Q_t$ & 0.90 & 0.1 & $-0.905$ & 0.93 & 0.2 & $-0.948$ & \w 0.38 & \w 0.0 & \w $-0.131$ & \w 0.39 & \w 0.7 & \w $-0.203$ & \w 0.55 & \w 24.5 & \w $-2.741$ & \w 0.55 & \w 25.2 & \w $-2.819$ \\
A2b: w/o load $Q_t$ & 0.88 & 0.1 & $-0.887$ & 0.93 & 0.3 & $-0.955$ & \w 0.38 & \w 0.0 & \w $-0.129$ & \w 0.39 & \w 0.3 & \w $-0.164$ & \w 0.55 & \w 24.5 & \w $-2.742$ & \w 0.55 & \w 25.0 & \w $-2.793$ \\
\midrule
A2a: w/o RRC state $\rho_t$, $\tau_t$ & 0.85 & 0.1 & $-0.860$ & 0.91 & 0.4 & $-0.938$ & 0.11 & 0.0 & 0.163 & 0.13 & 0.5 & 0.096 & \w 0.43 & \w 25.9 & \w $-2.760$ & \w 0.45 & \w 26.4 & \w $-2.830$ \\
\bottomrule
\end{tabular}
\end{table*}

Table~\ref{tab:ablation} groups the variants by the removed component. The task compute $c$ (A1) matters only on the weak link: at $-8$dB the policy keeps offloading OD and PE, and its reward falls by 0.07-0.10 in S2-S4, whereas at 0dB it stays within 0.01. The load $Q_t$ drives S3 and S4: without it (A2b) the policy runs locally on MAXN even when the server is idle, raising the normalized energy from 0.10 to 0.38 in S3 and from 0.44 to 0.55 in S4, as large a loss as removing $Q_t$ together with the RRC state (A2). The S4 violation rate stays within 23.6-26.4\% for all variants, so load observation is not needed for the violation gain; its main effect is lower energy. Without the RRC state (A2a), the S4 reward falls to $-2.760$ and $-2.830$ (full model $-2.538$ and $-2.565$) and the violation rate rises to 25.9\% and 26.4\%, while S2 and S3 change by at most 0.03. The main policy, trained without VLM-T, reached $-2.738$ in S4 on requests that include VLM-T, no lower than a policy trained with VLM-T (PPO-T, $-2.863$). LinUCB with the same state but no sequential credit ($\gamma = 0$, Table~\ref{tab:main}) matched the proposed policy in S1 and S2 but was 0.25 and 0.35 lower in S3 and S4; with $\alpha = 1$ it locked into local execution in S1 and S2 after one penalized offload (Proposition~1). With $\alpha = 20$ it avoids the lock-in, but the extra exploration lowers its reward in S3 and S4.

\subsection{Real-Device Validation}\label{sec:real}

\begin{table}[t]
\centering
\caption{Real-device validation on S4 ($-8$dB trace, medium severity, MEC-server load schedule): energy normalized to Always-Local (E) / latency violation rate in \% (V); mean over runs and, in the last column, over the three modes.}
\label{tab:real}
\scriptsize
\setlength{\tabcolsep}{3pt}
\begin{tabular}{lcccc}
\toprule
\textbf{Policy} & \textbf{MAXN} & \textbf{30W} & \textbf{15W} & \textbf{Mean} \\
\midrule
\multicolumn{5}{l}{\textit{VLM-S (local 3B / remote 7B)}} \\
Always-Local & 1.00 / 0.0 & 1.00 / 100 & 1.00 / 100 & 1.00 / 66.7 \\
Load-threshold rule & 0.77 / 1.6 & 0.56 / 50.1 & 0.38 / 46.4 & 0.57 / 32.7 \\
SW-UCB (RSSI)~\cite{r15} & 1.36 / 47.1 & 0.75 / 56.2 & 0.53 / 66.4 & 0.88 / 56.6 \\
\rowcolor{gray!18}\textbf{PPO (ours)} & \textbf{0.77 / 1.4} & \textbf{0.56 / 51.0} & \textbf{0.48 / 47.1} & \textbf{0.60 / 33.2} \\
\midrule
\multicolumn{5}{l}{\textit{VLM-T (local SmolVLM2 / remote 7B)}} \\
Always-Local & 1.00 / 0.0 & 1.00 / 85.4 & 1.00 / 100 & 1.00 / 61.8 \\
Load-threshold rule & 0.89 / 2.3 & 0.61 / 49.3 & 0.39 / 47.7 & 0.63 / 33.1 \\
SW-UCB (RSSI)~\cite{r15} & 1.53 / 43.6 & 0.79 / 54.5 & 0.52 / 58.0 & 0.95 / 52.0 \\
\rowcolor{gray!18}\textbf{PPO-T (ours)} & \textbf{0.88 / 1.4} & \textbf{0.69 / 46.7} & \textbf{0.48 / 50.4} & \textbf{0.68 / 32.8} \\
\bottomrule
\end{tabular}
\\[2pt]
\parbox{\columnwidth}{\scriptsize E is normalized per block to Always-Local with the same local model (Qwen-3B or SmolVLM2) under the same question schedule. PPO: seeds 0-3; PPO-T: seeds 0-4; other policies: five runs per mode.}
\end{table}

We replayed S4 on the AGX Orin: the medium-severity $-8$dB trajectory through the outage injection of Section~\ref{sec:factors}, and the load schedule through a load generator on the MEC server, for 10 minutes per run (Table~\ref{tab:real}). PPO ran seeds 0-3 (seed 4 excluded, as in Table~\ref{tab:main}), PPO-T seeds 0-4, and the other policies five runs each. Energy is the Jetson board energy integrated from INA3221 power over each question and normalized to Always-Local under the same question schedule, and the violation rate is over the questions each policy processed in the same 10 minutes. The 5G uplink latency during these runs was about 250-400ms, compared with about 100ms in our earlier runs with the same setup. Over the three modes PPO violated 33.2\% vs.\ 56.6\% for SW-UCB at an energy of 0.60 vs.\ 0.88. The gap was largest on MAXN (1.4\% vs.\ 47.1\%; 0.2\% vs.\ 36.0\% in simulation of the same $-8$dB, medium-severity condition), where PPO ran every question locally under load ($Q_t \ge 2$) at a lower long-question quality (0.37 vs.\ 0.42) and answered 129 questions per run vs.\ 90, while SW-UCB kept offloading into the queue and used more energy than Always-Local (1.36). On 30W and 15W the local 3B violates every latency constraint, so the gain came from load-free intervals and was smaller (51.0\% vs.\ 56.2\%, 47.1\% vs.\ 66.4\%). PPO-T followed the same pattern (1.4\% vs.\ 43.6\% on MAXN; 0.2\% vs.\ 34.0\% in simulation). The load-threshold rule matched PPO in every mode (32.7\% at 0.57; VLM-T 33.1\% at 0.63), but it needs a threshold for each task, mode, and local model (for VLM-S on MAXN, 30W, and 15W: 2, always remote, and 5), whereas one PPO policy covers all of them and had 1.4-3.8 points fewer violations at load levels absent from training (Section~\ref{sec:limits}).

\section{Discussion and Limitations}\label{sec:discussion}

\subsection{Discussion}\label{sec:disc}

Policies that observe only the link, SW-UCB and the field rule, kept offloading into the queue (40.4\% and 53.9\% S4 violations), whereas the proposed policy and a load-threshold rule on the same state stayed near 23.5\% in simulation and 33\% on the real device (Section~\ref{sec:s3s4} and Table~\ref{tab:real}). Learning adds a practical benefit rather than a lower violation rate: one policy covers four tasks and three power modes without per-combination thresholds and degrades less at load levels absent from training.

\subsection{Limitations}\label{sec:limits}

\textbf{Link and environment.} Weak signal was emulated by uplink outages along a field trajectory rather than by real mobility, so link-layer retransmission and handover are not captured. Reconnection occurred irregularly for idle times of 11-19s (rarely up to 10s, always at 20s), possibly because of background traffic such as NTP queries sent over 5G every 32s, which the simulator's timer model omits. Modem power was measured only for PE in one condition, without tail energy.

\textbf{Training and generalization.} The costs must be re-measured and the policy retrained when the hardware, models, or network change, and it inherits any sim-to-real gap. Training and evaluation trajectories come from the same three drives; when one drive was held out of training and used for evaluation, the reward stayed within 0.05 and the violation rate within 0.6 points of the policy trained on all three. With two or four VLM server slots instead of one, the policy overestimates the queueing delay and has more violations than the load-threshold rule, so it must be retrained when the server capacity changes. The quality $\hat{g}$ is a per-model, per-type mean that ignores per-answer correctness.

\textbf{Load observation.} The policy observed $Q_t$ exactly in all experiments, including the real-device runs, so the results assume an ideal load report. In a separate simulation run, a 10s delay in $Q_t$ raised the S4 violation rate only from 23.5\% to 24.4\%, but using only the value carried by the last offload response kept the policy local after the load ended, raising the rate to 27.8\% and the energy from 0.44 to 0.57, still below SW-UCB in the same run (40.4\%); a lightweight periodic load report is therefore needed. Uniform $\pm 1$ or Poisson noise in the reported $Q_t$ raises the S4 violation rate from 23.5\% to 25.1-25.5\% (load-threshold rule 25.9-26.3\%), and $\pm 2$ noise to 29.2\% (26.2\%). At load values absent from training ($Q_t \in \{1, 3, 4\}$), the policy has 1.4-3.8 points fewer violations than the load-threshold rule, which only separates the trained levels.

\subsection{Future Work}\label{sec:future}

We plan to measure modem power during movement, to add multiple servers and server models, and to study several robots sharing one server, as in a micro data center (MDC) built on Korean AI SoCs.

\section{Conclusion}\label{sec:conclusion}

This paper addressed when a robot should offload vision-language model (VLM) inference to a private 5G MEC. Real-testbed measurements of four tasks in three power modes showed that model compute sets the offloading gain and sorts requests into three regimes without server load. Link degradation mattered only in the state-dependent regime, server load even for the small VLM, and the RRC state depended on previous choices. We formulated a constrained MDP whose fixed violation penalty acts as a Lagrange multiplier, showed that a myopic policy without exploration can lock into local execution under RRC dynamics, and trained a PPO-based congestion-aware sequential offloading policy that observes these factors. In simulated field drives, it cut the server-load violation rate from 40.4\% (SW-UCB) to 23.6\%; observing the load mainly saved energy, and the RRC state reduced violations. On the real device in MAXN, where the local model meets the constraint, it cut the rate from 47.1\% to 1.4\% (mean over four seeds), at slightly lower long-question quality. It also served an unseen tiny VLM sharing the remote model without retraining. In S1, its reward was slightly below those of Always-Remote and SW-UCB, and because it assumed exact load observation, deployment needs a lightweight periodic load report. Offloading VLMs to shared edge servers thus requires observing the model compute, power mode, server load, and modem state together with the link.

\vspace{12pt}

\end{document}